\documentclass[aps,pra,twocolumn,superscriptaddress,reprint]{revtex4-1}

\usepackage{float}
\usepackage{graphicx}
\usepackage{amsmath}
\usepackage{bm}
\usepackage{color}
\usepackage{amssymb}
\usepackage{mathrsfs}
\usepackage{dcolumn}
\usepackage{multirow}
\usepackage[colorlinks=true,linkcolor=blue,urlcolor=blue,citecolor=blue]{hyperref}

\begin{document}
	
	\title{Crystalline splitting of $d$ orbitals in two-dimensional regular optical lattices}
	
	\author{Hua Chen}
	\email{Electronic address: hwachanphy@zjnu.edu.cn} 
	\affiliation{Department of Physics, Zhejiang Normal University, Jinhua 321004, China}
	
	\author{X. C. Xie}
	\affiliation{International Center for Quantum Materials, School of Physics, Peking University, Beijing 100871, China}
	\affiliation{Collaborative Innovation Center of Quantum Matter, Beijing 100871, China}
	\affiliation{CAS Center for Excellence in Topological Quantum Computation, University of Chinese Academy of Sciences, Beijing 100190, China}
	
	\begin{abstract}
		In solids, crystal field splitting refers to 
		the lifting of atomic orbital degeneracy by the surrounding ions through the static electric field.
		Similarly, we show that the degenerated $d$ orbitals, which were derived in the harmonic oscillator approximation, 
		are split into a low-lying $d_{x^2+y^2}$ singlet and a $d_{x^2-y^2/xy}$ doublet 
		by the high-order Taylor polynomials of triangular optical potential.
		The low-energy effective theory of the orbital Mott insulator at $2/3$ filling is generically described by the Heisenberg-Compass model, 
		where the antiferro-orbital exchange interactions of compass type 
		depend on the bond orientation and are geometrically frustrated in the triangular lattice.  
		While, for the square optical lattice, the degenerated $d$ orbitals are split into a different multiplet structure, {\it i.e.} a low-lying $d_{x^2\pm y^2}$ doublet and a $d_{xy}$ singlet, 
		which has its physical origin in the $C_{4v}$ point group symmetry of square optical potential.
		Our results build a bridge between ultracold atom systems and solid-state systems for the investigation of $d$-orbital physics.  
	\end{abstract}
	
	\date{\today}
	
	\maketitle
	
	\section{Introduction}

	In transition metal oxides, the degenerated $d$ orbitals are split into a set of orbital multiplets,
	typically a $t_{2g}$ triplet and a $e_{g}$ doublet for the cubic perovskite structure, 
	by the surrounding oxygen anions through the crystalline electric field, 
	accompanied by the breaking of the full spherical symmetry of a free atom~\cite{Tokura00,Maekawa04}.
	Hence, the key feature of $d$ orbitals in solids is 
	that both the orbital degeneracy and orientational anisotropy 
	are governed by the finite point group symmetry of solids. 
	The crystal structure is reflected in the orbital multiplets and is the origin of various interesting phenomena,
	covering metal-insulator transitions~\cite{Imada98}, superconductivity~\cite{Damascelli03,Mackenzie03,Armitage10,Stewart11,Dai15}, 
	and colossal magneto-resistance~\cite{Dagotto01,Khaliulli05,Oles17}.
	More recently, the forefront of experimental research has focused on the Kitaev material $\alpha$-RuCl$_3$, 
	in which the relativistic pseudospin-$1/2$ states arise from the delicate balance of the crystalline electric field, spin-orbit coupling, and strong correlation~\cite{Jackeli09,Chaloupka13}.
	This material exhibits strongly anisotropic pseudospin exchange interactions originated from the bond-directional nature of $d$ orbitals via spin-orbital entanglement, 
	and shows the increasing experimental evidence in supporting the celebrated Kitaev spin-liquid physics
	~\cite{Banerjee16,Banerjee17,Do17,Jansa18,Kasahara18}.

	Ultracold atom gases offer highly controllable platforms 
	for the quantum simulations of artificial solids in optical lattices,
	which have served successfully as a complementary set up to solid-state systems during the past decade~\cite{Bloch12}. 
	As a paradigmatic example, the $p$-orbital physics in optical lattices 
	attracts intensive research interests for the orbital degree of freedom~\cite{Lewenstein11,Li16,Kock16}.
	Interesting many-body phenomena were predicted including
	unconventional Bose-Einstein condensation~\cite{Isacsson05,Liu06,Kuklov06}, 
	supersolid phase~\cite{Scarola05}, stripe ordering~\cite{Wu06}, Wigner crystallization~\cite{Wu07},
	and orbital ordering in Mott insulators~\cite{Liu08,Wu08,Pinheiro13}.
	Importantly, the chiral $p_x \pm ip_y$ superfluidity has been successfully observed in recent experiments~\cite{Wirth11,Olschlager13,Kock15}.
	However, the $p$ orbitals are essentially different from 
	the $d$ orbitals for both orbital degeneracy and orientational anisotropy.
	Particularly exciting is the recent experimental advance 
	in the observation of $d$ orbitals in optical lattices~\cite{Hemmerich11,Hemmerich12,Zhai13,Zhou16},
	which makes an important step towards genuinely emulating $d$-orbital physics of solid-state systems. 
	Here we report that the degeneracy of $d$ orbitals, 
	which was predicted in the harmonic oscillator (HO) approximation,
	is partly removed by the high-order Taylor polynomials (HOTPs) of optical potential 
	in both triangular and square optical lattices.
	In the triangular lattice, the orbital Mott insulator is further studied
	based on the remaining degeneracy between $d_{x^2-y^2}$ and $d_{xy}$ orbitals. 
	The corresponding orbital exchange Hamiltonian is generically described by the Heisenberg-compass model,
	where the anisotropic compass interactions have roots in the orbital orientational anisotropy and are geometrically frustrated. For the square lattice, in particular, 
	we have derived a selection rule on the orbital angular momentum, 
	and show that the geometry of square optical lattice plays a crucial role in determining the orbital multiplets.

	\section{Triangular optical lattice}	
	
	\begin{figure}
		\centering
		\includegraphics[width=0.48\textwidth]{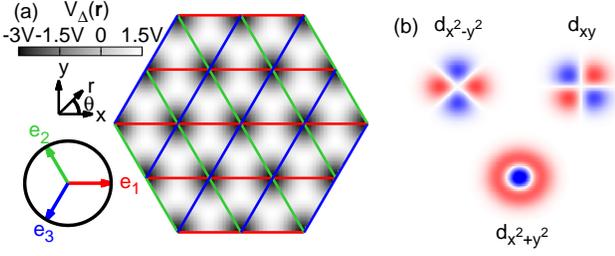}	
		\caption{(color online). (a) Grey map of the triangular optical potential $V_\triangle\left(\bm{r}\right)$.
			$\{\bm{e}_1,\bm{e}_2,\bm{e}_3\}$ are the bond vectors of triangular lattice.
			(b) Structure of partially lifted degeneracy of $d$-orbital multiplets in the triangular optical lattice.}
		\label{fig:triangle}
	\end{figure}
	
	The triangular optical potential has been theoretically proposed~\cite{Petsas94,Jaksch05} and experimentally realized~\cite{Becker10,Struck11,Struck13} 
	using three linearly polarized laser beams.
	It is mathematically described by 
	$V_\triangle\left(\bm{r}\right) \equiv -V\sum_{i=1}^{3}\cos\left(\bm{b}_i \cdot \bm{r} \right)$,
	where the reciprocal lattice vectors 
	$\bm{b}_1=\frac{2\pi}{a}\left(\hat{x}+\frac{1}{\sqrt{3}}\hat{y}\right)$,
	$\bm{b}_2=\frac{2\pi}{a}\left(-\hat{x}+\frac{1}{\sqrt{3}}\hat{y}\right)$
	and $\bm{b}_3=-\frac{4\pi}{\sqrt{3}a}\hat{y}$
	with $a$ the lattice spacing.
	Figure~\ref{fig:triangle}(a) plots the periodic landscape of optical potential $V_\triangle\left(\bm{r}\right)$, 
	the spatial modulation of which realizes the triangular lattice.
	Since the lattice is invariant under primitive translations of bond vectors \{$\bm{e}_1$,$\bm{e}_2$,$\bm{e}_3$\}, 
	we will focus on the lattice site at the origin of coordinates to simplify the discussion.
	Switching to polar coordinates $\left(r,\theta\right)$, the optical potential can be expressed in terms of Bessel functions of the first kind
	via the Jacobi-Anger expansion,
	\begin{equation}
	V_\triangle\left(\bm{r}\right) = \sum_{{\ell}=-\infty}^{+\infty}V_\triangle^{\ell}\left(r\right)\exp\left[6i{\ell}\theta\right],
	V_\triangle^{\ell}\left(r\right)\equiv-3VJ_{6\ell}\left(\frac{4\bar{r}}{\sqrt{3}}\right)
	\label{eq:triexp}
	\end{equation}
	with the dimensionless radial distance $\bar{r}\equiv \pi r/a$.
	A Taylor series expansion of the isotropic component 
	$V_\triangle^{\ell=0}=-3V+4V\bar{r}^2+\mathcal{O}\left(\bar{r}^4\right)$ 
	in Eq.~(\ref{eq:triexp}) yields a 2D harmonic trapping of frequency 
	$\omega=\sqrt{8\pi^2V/Ma^2}$ ($M$ is the mass of trapped atoms).
	In the deep lattice limit, the Wannier functions in the optical potential $V_\triangle\left(\bm{r}\right)$  
	are well approximated by the corresponding eigenfunctions of HO~\cite{Isacsson05,Liu06}.
	Due to the isotropic nature of the 2D HO, 
	the eigenfunctions have simultaneous eigenstates with the $z$-axis angular momentum operator $L_z=-i\hbar\partial_\theta$
	and thus can be written in the axial states
	\begin{equation}
	\Psi_{\left[n,m\right]}\left(\bm{r}\right) \equiv R_{\left[n,m\right]}\left(r\right)\exp\left[im\theta\right],\nonumber
	\end{equation}
	with $n$ and $m$ labeling the quanta of the 2D HO  
	and $z$-axis angular momentum, respectively (see Appendix~\ref{app:HO} for details). 
	The explicit forms of eigenfunctions $\Psi_{\left[n,m\right]}\left(\bm{r}\right)$ for $n=2$, which we will refer to as $d$ orbitals hereafter, are listed in Table~\ref{tab:eigfun}.
	
	\begin{table}
		\caption{\label{tab:eigfun}
			The $d$-orbital wave functions $\Psi_{\left[n=2,m\right]}\left(\bm{r}\right)$ of the 2D isotropic harmonic oscillator of frequency $\omega$ with $\beta\equiv\sqrt{M\omega/\hbar}$.} 
		\begin{ruledtabular}
			\begin{tabular}{ccc}
				$n$	&	$m$	& 	$\Psi_{\left[n,m\right]}\left(\bm{r}\right)\equiv R_{\left[n,m\right]}\left(r\right)\exp\left[im\theta\right]$	\\
				\hline
				\multirow{3}{*}{}  &	$+2$	&		$\Psi_{\left[2,+2\right]}\left(\bm{r}\right)=\frac{\beta^3}{\sqrt{2\pi}}r^2\exp\left[-\frac{\beta^2 r^2}{2}\right]\exp\left[+2i\theta\right]$\\
				$2$ 				  &	$0$	&	
				$\Psi_{\left[2,0\right]}\left(\bm{r}\right)=\frac{\beta}{\sqrt{\pi}}\left[\left(\beta r\right)^2-1\right]\exp\left[-\frac{\beta^2 r^2}{2}\right]$\\		
				&	$-2$	&
				$\Psi_{\left[2,-2\right]}\left(\bm{r}\right)=\frac{\beta^3}{\sqrt{2\pi}}r^2\exp\left[-\frac{\beta^2 r^2}{2}\right]\exp\left[-2i\theta\right]$\\			  						
				
			\end{tabular}
		\end{ruledtabular}
	\end{table}
	
	Next, we will show that the high-order polynomials in the Taylor series expansion of 
	isotropic potential $V_\triangle^{\ell=0}\left(r\right)$ will further lift the degeneracy of $d$-orbital complex. 
	To proceed, we expand field operators in the $d$-orbital Wannier basis 
	and obtain the second quantization form of HOTPs in $V_\triangle\left(\bm{r}\right)$ in Eq.~(\ref{eq:triexp}) 
	\begin{equation}
	\mathscr{H}_\triangle =  \sum_{m_1 m_2} \sum_{\ell=-\infty}^{+\infty}
	\left\langle \Psi_{\left[2,m_1\right]} | \triangle^\ell | \Psi_{\left[2,m_2\right]} \right\rangle
	\hat{\Psi}_{\left[2,m_1\right]}^\dagger \hat{\Psi}_{\left[2,m_2\right]},
	\label{eq:cefm}
	\end{equation}
	where the HOTPs
	$\triangle^\ell\left(\bm{r}\right) \equiv V_\triangle^\ell\left(r\right)\exp\left[6i\ell\theta\right]
	+\left(3V-4V\bar{r}^2\right)\delta_{\ell,0}$
	and $\hat{\Psi}_{\left[2,m\right]}^\dagger$ ($\hat{\Psi}_{\left[2,m\right]}$) 
	creates (annihilates) an atom in the state $\Psi_{\left[n=2,m\right]}$.
	It is easy  to verify that the matrix elements of anisotropic potential  
	$\triangle_{m_1m_2}^{\ell\ne 0}\equiv\left\langle \Psi_{\left[2,m_1\right]} | \triangle^\ell | \Psi_{\left[2,m_2\right]} \right\rangle$ have no contributions because of the vanishing integrals of azimuthal parts over polar angle $\theta$. 
	While, for the isotropic case $\ell=0$, the matrix $\triangle_{m_1m_2}^{\ell=0}$ has nonvanishing diagonal elements
	\begin{eqnarray}
	\{\triangle_{\pm2,\pm2}^{\ell=0},\triangle_{0,0}^{\ell=0}\}=
	&-&\frac{E_\text{R}}{12}\sum_{l=0}^{\infty}
	\left(-\frac{1}{3}\sqrt{\frac{E_\text{R}}{2V}}\right)^l\frac{1}{\left(l+2\right)!} \nonumber \\
	&\times& \{l^2+7l+12,2l^2+10l+14\}\nonumber
	\end{eqnarray}
	with the recoil energy $E_\text{R}\equiv 4\hbar^2\pi^2/Ma^2$.
	The axial states $\Psi_{\left[n=2,m=\pm2\right]}$ have the identical correction 
	on their energy levels by the HOTPs
	$\triangle^{\ell=0}\left(\bm{r}\right)$. 
	The reason can be traced back to the fact that their eigenfunctions share the same radial function, as listed in Table~\ref{tab:eigfun}.
	A unitary transformation 
	$\Psi_{\left[n=2,m=\pm2\right]}\equiv \left(d_{x^2-y^2}\pm id_{xy}\right)/\sqrt{2}$ 
	and $\Psi_{\left[n=2,m=0\right]}\equiv d_{x^2+y^2}$~\cite{dx2y2},
	followed by an irrelevant energy shift of $\triangle_{0,0}^{\ell=0}$,
	cast $\mathscr{H}_\triangle$ in Eq.~(\ref{eq:cefm}) into a concrete form
	\begin{equation}
	\mathscr{H}_\triangle = \triangle \left(d_{x^2-y^2}^\dagger d_{x^2-y^2} + d_{xy}^\dagger d_{xy}\right)
	\label{eq:cef}
	\end{equation}
	with $\triangle\equiv\triangle_{\pm2,\pm2}^{\ell=0}-\triangle_{0,0}^{\ell=0}
	=\frac{E_\text{R}}{12}\exp\left[-\frac{1}{3}\sqrt{\frac{E_\text{R}}{V}}\right]$ 
	describing the energy splitting between $d_{x^2-y^2/xy}$ and $d_{x^2-y^2}$ orbitals.
	In the deep lattice limit, $V \gg E_\text{R}$, the energy splitting $\triangle$ saturates at $E_\text{R}/12$, 
	and the $d$-orbital complex is well separated from the $s$ and $p_{x,y}$ orbitals in energy, 
	primarily by the HO frequency $\hbar\omega = \sqrt{2VE_\text{R}}$, 
	indicating the validity of first-order perturbation treatment above.  
	As is summarized in Fig.~\ref{fig:triangle}(b), 
	the $d$-orbital complex splits into a low-lying $d_{x^2+y^2}$ singlet and a $d_{x^2-y^2/xy}$ doublet, 
	which is analogous to the crystalline electric field splitting in solid-state physics~\cite{Pavarini12}. 
	When a $d$-orbital ion is embedded in a solid, 
	the full fivefold degeneracy of hydrogen-like $d$ orbitals, which is protected by the spherical symmetry of a free atom,
	is lifted by the charged neighboring ions through the crystal field potential.
	While, the splitting of $d$-orbital complex in the triangular optical lattice is rooted in the different radial functions
	between $d_{x^2-y^2/xy}$ and $d_{x^2-y^2}$ orbitals 
	through the isotropic high-order optical potential $\triangle^{\ell=0}\left(\bm{r}\right)$.
	We will show that the anisotropic optical potential can also contribute to 
	the degeneracy lifting in a different manner, 
	see discussions on the square optical lattice latter.  
	
	\begin{figure}
		\centering
		\includegraphics[width=0.48\textwidth]{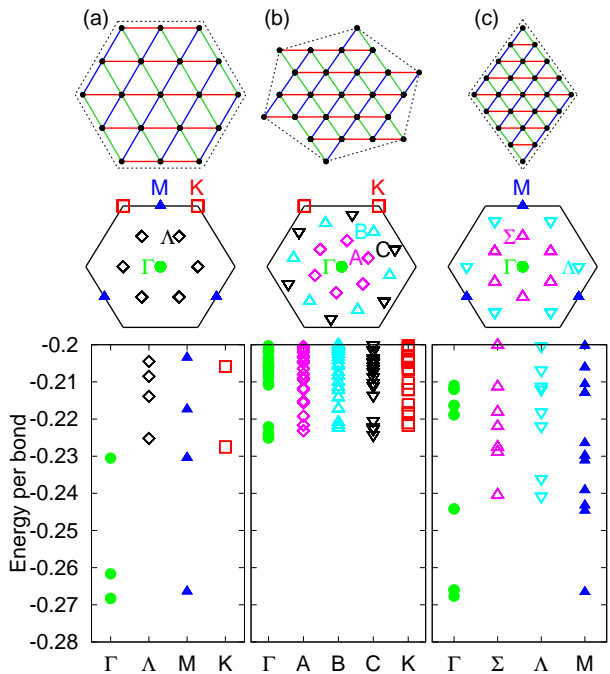}	
		\caption{ (color online). 
			Low-energy spectra (bottom panel) of quantum $120^\circ$ compass model on the finite-size clusters of (a) 12 sites, (b) 21 sites, and (c) 16 sites
			with exchange couplings $\left(J_\text{C},J_\text{H}\right)=\left(1,0\right)$. 
			The $x$ axis labels the momenta of many-particle states, which are marked in the hexagonal Brillouin zone (middle panel). 
			The corresponding samples of finite-size clusters with periodic boundary conditions (black dashed lines) are shown in the top panel. }
		\label{fig:ed}
	\end{figure}
	
	It is then interesting to explore the interplay between the geometrical frustration of triangular lattice 
	and the quantum fluctuation, which is enhanced by the remaining degeneracy of $d_{x^2-y^2}$ and $d_{xy}$ orbitals.
	The pioneering works have studied $p_{x,y}$-orbital Mott insulators with spinless fermions 
	and found various exotic orbital orderings in the classical ground states~\cite{Liu08,Wu08}.
	To this end, it is necessary to carry out a strong coupling study of the correlated $d$-orbital systems.
	Let us start with the case that spinless fermions interact with each other through a a general central potential $\hat{U}\left(r\right)$.
	The interacting Hamiltonian is constructed in terms of the Haldane pseudopotentials
	\begin{equation}
	\mathscr{H}_\text{I}=\sum_m \sum_{i<j} v_m \mathcal{P}_m\left(ij\right) \nonumber
	\end{equation}
	where $\mathcal{P}_m\left(ij\right)$ is the projection operator 
	which selects out states in which particles $i$ and $j$ have relative angular momentum $m$~\cite{Haldane90}. 
	According to the Fermi (Bose) statistics, the many-particle state of fermions (bosons) 
	should be antisymmetric (symmetric) upon interchanging two particles, which requires that $m$ is odd (even).
	Thus, the pseudopotential set $\left\{v_m\right\}$ with odd $m$ provide a complete and unique description of interaction $\hat{U}\left(r\right)$ for spinless fermions.
	For a short-range interaction $\hat{U}\left(r\right)$, 
	the leading interaction between $d$ orbitals is described by 
	\begin{equation}
	\mathscr{H}_\text{I} = U\left[\left(\hat{n}_{x^2-y^2}+\hat{n}_{xy}\right)\hat{n}_{x^2+y^2}
	+2\hat{n}_{x^2-y^2}\hat{n}_{xy}\right],
	\label{eq:int}
	\end{equation}
	where $U\equiv 3v/16\pi$ and the Haldane pseudopotentials $v_{\pm1}\equiv v$ 
	are the short-range components of $\hat{U}\left(r\right)$ in active channels $m=\pm1$ (see Appendix~\ref{app:Haldane} for details).
	The interactions between the $d$ orbitals and the low-lying $s$ and $p_{x,y}$ orbitals 
	cannot lift the remaining degeneracy of $d$ orbitals in Eq.~(\ref{eq:cef}),  
	which is protected by the continuous rotation symmetry.
	The well separated $s$ and $p_{x,y}$ orbitals are reminiscent of the closed shells in solid-state systems
	and remain inactive at low energy scales.
	Interestingly, the $d$ orbitals can be prepared by the direct transfer between even-parity orbitals $s\to d$ with the fidelities as high as $97$-$99\%$ in the recent experiments~\cite{Zhai13,Zhou16}.
	Therefore, in the following, we shall only consider the interaction between $d$ orbitals.
	For the case that the $d$ orbitals are partially occupied by $n$ spinless fermions, we will refer to it as $d^n$ configuration. Including the crystalline splitting $\mathscr{H}_\triangle$ in Eq.~(\ref{eq:cef}) 
	and the on-site interaction $\mathscr{H}_\text{I}$ in Eq.~(\ref{eq:int}),
	the ground state of $d^2$ configuration is an orbital doublet 
	with one fermion occupying the low-lying $d_{x^2+y^2}$ orbital 
	and the other one occupying either $d_{x^2-y^2}$ or $d_{xy}$ orbital, 
	and simply inherits the partially degeneracy of $d$-orbital complex.
	It is convenient for later discussions to define the pseudospin operators  
	$\{\tau^+,\tau^-\} \equiv \{d^\dagger_{x^2-y^2}d_{xy}\hat{n}_{x^2+y^2},d^\dagger_{xy}d_{x^2-y^2}\hat{n}_{x^2+y^2}\}$,
	which flip the states of orbital doublet. 
	The $z$ component of pseudospin $\bm{\tau}$-vector follows through the spin-$1/2$ angular momentum algebra $\tau^z=\left[\tau^+,\tau^-\right]$. 
	In the strongly correlated regime, orbital fluctuation is the remaining low energy degree of freedom.
	Therefore, the effective model is captured by the orbital superexchange interactions 
	between sites $i$ and $j$, which arise from 
	the virtual charge excitations $(d^2)_i(d^2)_j\rightleftharpoons (d^3)_i(d^1)_j$ 
	through the hopping process $t_{\mu\nu}d^\dagger_{i\mu}d_{j\nu}$ ($\mu,\nu=x^2-y^2,xy,x^2+y^2$).
	Employing the second-order perturbation theory in Ref.~\cite{Kuklov03}, 
	we derive the effective Hamiltonian in Appendix~\ref{app:Heff}.
	It is generically described by the Heisenberg-Compass model 
	$\mathscr{H}^\text{eff}_\triangle = \mathscr{H}_\text{H}+\mathscr{H}_\triangle^{\text{120}^\circ}$,
	where the isotropic Heisenberg term 
	$\mathscr{H}_\text{H}=J_\text{H}\sum_{i\eta\gamma}
	\bm{\tau}_i\cdot\bm{\tau}_{i+\eta\bm{e}_\gamma}$ 
	and the anisotropic compass term~\cite{Brink04,Nussinov15}
	\begin{eqnarray}
	\mathscr{H}_\triangle^{\text{120}^\circ}=J_\text{C}\sum_{ i\gamma\eta }
	\tau_i^\gamma \tau_{i+\eta\bm{e}_\gamma}^\gamma
	\label{eq:compass}
	\end{eqnarray} 
	with
	\begin{eqnarray}
	&\tau^\gamma=\tau^z\cos\left[4\theta_\gamma\right]+\tau^x\sin\left[4\theta_\gamma\right],
	\bm{e}_\gamma=\hat{x}\cos\theta_\gamma+\hat{y}\sin\theta_\gamma,& \nonumber \\
	&\{\theta_1,\theta_2,\theta_3\}=\{0,\frac{2\pi}{3},\frac{4\pi}{3}\},\eta=\pm 1.& \nonumber 
	\end{eqnarray} 
	The superexchange couplings are given by 
	\begin{equation}
	\{J_\text{H},J_\text{C}\}=\{t_{\pi}t_{\sigma}/U,\left(t_{\sigma}-t_{\pi}\right)^2/2U\} \nonumber
	\end{equation}
	with $t_{\pi}$ ($t_{\sigma}$) denoting the intra-orbital $\pi$($\sigma$)-bonding state 
	of $d_{xy}$ ($d_{x^2-y^2}$) orbital. 
	It is worth noting that the $\pi$-bonding axis lies in the nodal plane of $d_{xy}$ orbital.
	As a result, the $\pi$ bonding is typically much weaker than the $\sigma$ bonding,
	and the corresponding antiferro-orbital compass interaction dominates over the ferro-orbital Heisenberg interaction 
	($J_\text{H}<0$ is due to the opposite sign of $t_\pi$ and $t_\sigma$).
	This is reminiscent of the Heisenberg-Kitaev model in the afore-mentioned Kitaev material $\alpha$-RuCl$_3$ 
	with the dominant Kitaev coupling~\cite{Jackeli09,Chaloupka13}.
	Solving the quantum Heisenberg-compass model remains a challenging problem.
	Nevertheless, it is instructive to first determine the ground state of dominant part, ${\it i.e.}$ quantum compass model~\cite{Nussinov15}, 
	for understanding the phase diagram of quantum Heisenberg-Compass model. 
	The particularity of quantum compass model $\mathscr{H}_\triangle^{\text{120}^\circ}$ in Eq.~(\ref{eq:compass}) is that 
	along the bond vector $\pm \bm{e}_\gamma$ $\left(\gamma=1,2,3\right)$ 
	the exchange interaction involves the pseudospin $\tau^{\gamma}$ of two sites connected by the bond,
	and the pseudospin components $\tau^{1,2,3}$ intersect in the $zx$-plane at an effective angle of $120^\circ$. 
	The quantum $120^\circ$ model is first introduced as an effective model for perovskite $e_g$ orbital systems~\cite{Brink99},
	which is closely related to the well-known quantum compass model~\cite{Kugel82}.
	Apparently, it is impossible to minimize the antiferro-orbital interactions for all three bonds 
	on an elementary triangle simultaneously due to the geometrical frustration.
	In this case, exotic quantum states are usually promoted by the geometrical frustration via spontaneous symmetry breaking.
	To capture the quantum fluctuations, we resort to Lanczos exact diagonalization on finite-size clusters.
	As illustrated in Figs.~\ref{fig:ed} (a) and \ref{fig:ed}(b), 
	we first employ the clusters with $60^\circ$ equilateral parallelograms to avoid the cluster shape dependence of results~\cite{Marland79}.
	The corresponding energy spectra are carefully analyzed by
	extracting the momentum of each eigenstate.
	One key signature in the spectrum of 12-site cluster is that 
	several low-lying states are well separated from the excited states by a clear gap.   
	The energies of these low-lying states are much lower than the ground-state energy of 21-site cluster.
	It is well accepted that the quantum counterpart of classical ground state is a coherent superposition of low-lying eigenstates, 
	which are dubbed as quasidegenerate joint states (QDJSs)~\cite{Bernu92,Bernu94}.
	As shown in Fig.~\ref{fig:ed} (c), further studies on the 16-site cluster confirm that the energy spread of QDJSs decreases upon increasing the size of cluster.  
	Importantly, the QDJSs involve three degenerate states at the $M$ points of hexagonal Brillouin zone,
	which provides a strong evidence that the macroscopic symmetry-breaking state is of columnar type.
	Interestingly, the energies of QDJSs are close to the energy of classical columnar state, $-0.25J_\text{C}$ per bond.
	This classical state is also proposed as the ground state of $p_{x,y}$-orbital Mott insulators in Ref.~\cite{Wu08}.
	While, in the Heisenberg limit $\left(J_\text{H}<0,J_\text{C}=0\right)$, 
	the ferro-orbital exchange favors parallel alignments of nearest neighbor orbitals along bonds and is thus free of geometrical frustration.
	The transition between classical columnar phase and ferro-orbital phase occurs at the critical value $J_\text{C}=-8J_\text{H}/3$, above which the classical columnar state is stabilized. 
	As shown in Fig.~\ref{fig:ed}, the columnar phase is associated with the QDJSs at the $\Gamma$ and $M$ points of the hexagonal Brillouin zone.
	The interference between QDJSs at the $\Gamma$ and $M$ points breaks both the translation symmetry of triangular lattice
	and the point group symmetry from $C_6$ down to $C_2$ symmetry, which can be distinguished from the ferro-orbital phase.
	Experimentally, the symmetry breaking can be in principle detected by the time-of-flight interference~\cite{Bloch08}.  
	It is also noteworthy that the breaking of translation symmetry leads to the enlarged unit cell in the columnar phase.  
	In the time-of-flight noise correlation spectra, the momentum resolved interference spots will be observed
	at the corresponding reciprocal lattice points in the columnar phase,	from which the broken symmetries can be easily identified.

	\section{Square optical lattice}	
	
	\begin{figure}
		\centering
		\includegraphics[width=0.48\textwidth]{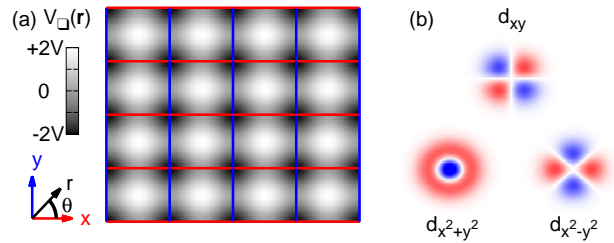}	
		\caption{(color online). (a) Grey map of the square optical potential $V_\square\left(\bm{r}\right)$.
			(b) Structure of partially lifted degeneracy of $d$-orbital multiplets in the square optical lattice.}
		\label{fig:square}
	\end{figure}
	
	Next, we turn to the square optical potential
	$V_\square \left(\bm{r}\right) = -V
	\left[
	\cos\left(\bm{b}_1 \cdot \bm{r}\right)
	+\cos\left(\bm{b}_2 \cdot \bm{r}\right)
	\right]$
	with the reciprocal lattice vectors 
	$\bm{b}_1=\frac{2\pi}{a}\hat{x}$ and $\bm{b}_2=\frac{2\pi}{a}\hat{y}$.
	The Jacobi-Anger expansion of square optical potential leads to
	\begin{equation}
	V_\square\left(\bm{r}\right) = \sum_{{\ell}=-\infty}^{+\infty}V_\square^{\ell}\left(r\right)\exp\left[4i{\ell}\theta\right],
	V_\square^{\ell}\left(r\right)\equiv-2VJ_{4\ell}\left(\frac{2\pi r}{a}\right).
	\label{eq:squareexp}
	\end{equation}
	The curvature at the bottom of isotropic component $V^{\ell=0}_\square=-2V+2V\pi^2r^2/a^2+\mathcal{O}\left(r^4\right)$
	in Eq.~(\ref{eq:squareexp}) dictates the HO frequency $\omega=\sqrt{4V\pi^2/Ma^2}$.
	The high-order correction on $d$-orbital complex is then described by 
	\begin{equation}
	\mathscr{H}_\square =  \sum_{m_1 m_2} \sum_{\ell=-\infty}^{+\infty}
	\left\langle \Psi_{\left[2,m_1\right]} | \square^\ell | \Psi_{\left[2,m_2\right]} \right\rangle
	\hat{\Psi}_{\left[2,m_1\right]}^\dagger \hat{\Psi}_{\left[2,m_2\right]},
	\label{eq:cefm-square}
	\end{equation}
	where  
	$\square^\ell\left(\bm{r}\right) \equiv V_\square^\ell\left(r\right)\exp\left[4i\ell\theta\right]
	+\left(2V-2V\pi^2r^2/a^2\right)\delta_{\ell,0}$.
	The nonzero diagonal elements in the isotropic channel $\ell=0$ are given by
	\begin{eqnarray}
	\{\square_{\pm2,\pm2}^{\ell=0},\square_{0,0}^{\ell=0}\}=
	&-&\frac{E_\text{R}}{16}\sum_{l=0}^{\infty}
	\left(-\frac{1}{4}\sqrt{\frac{E_\text{R}}{V}}\right)^l\frac{1}{\left(l+2\right)!} \nonumber \\
	&\times& \{l^2+7l+12,2l^2+10l+14\}\nonumber
	\end{eqnarray}
	While, for the anisotropic channel $\ell\ne0$, the integral over polar angle $\theta$ yields 
	a selection rule 
	$m_1-m_2=4\ell$,
	which has an intuitive meaning from the view of angular momentum conservation:
	$m_1$ ($m_2$) is the angular momentum in the final (initial) state and $4\ell$ is supplied by the square optical lattice 
	because it has a fourfold discrete rotational symmetry.
	The nonvanishing terms, satisfying the selection rule, are explicitly evaluated as   
	\begin{equation}
	\square^{\ell=1}_{+2,-2}=\square^{\ell=-1}_{-2,+2}
	=\frac{E_\text{R}}{16}\exp\left[-\frac{1}{4}\sqrt{\frac{E_R}{V}}\right]. \nonumber
	\end{equation}
	The reduction of continuous $z$-axis rotation symmetry lifts 
	the degeneracy of time-reversal partners $\Psi_{\left[n=2,m=\pm 2\right]}$ and quenches the orbital momentum.  
	Finally, a little algebra, together with an overall energy shift of $\square^{\ell=0}_{0,0}$, casts $\mathscr{H}_\square$ in Eq.~(\ref{eq:cefm-square}) into the form
	\begin{equation}
	\mathscr{H}_\square=\square\left(d^\dagger_{xy}d_{xy}-d^\dagger_{x^2-y^2}d_{x^2-y^2}-d^\dagger_{x^2+y^2}d_{x^2+y^2}\right)
	\nonumber
	\label{eq:cef-square}
	\end{equation}
	with $\square\equiv\frac{E_\text{R}}{16}\exp\left[-\frac{1}{4}\sqrt{\frac{E_R}{V}}\right]$ 
	describing the energy splitting between $d_{x^2\pm y^2}$ and $d_{xy}$ orbitals.
	Figure~\ref{fig:square}(b) depicts the structure of $d$-orbital multiplets in the square optical lattice.
	From symmetry aspects, the $\{d_{x^2+y^2},d_{x^2-y^2},d_{xy}\}$ orbitals 
	belong to the irreducible representations $\{A_1,B_1,B_2\}$ of $C_{4v}$ point group symmetry, respectively~\cite{Dresselhaus08}.
	It is noteworthy that the $C_{4v}$ symmetry is not sufficient to guarantee the degeneracy of $d_{x^2\pm y^2}$ doublet,
	which can be lifted in a checkerboard optical potential.
	
	In the $d^1$ configuration, the ground state is an orbit doublet 
	with one fermion occupying either $d_{x^2+y^2}$ or $d_{x^2-y^2}$ orbital. 
	In the large-$U$ limit, we next briefly discuss the corresponding low-energy effective model 
	that is constructed based on the ground-state doublet through the virtual charge excitations 
	$\left(d^1\right)_i\left(d^1\right)_j\rightleftharpoons (d^2)_i(d^0)_j$.
	For the case that the hopping integrals $t_{\mu\nu}$ is comparable to the crystalline splitting $\square$, 
	the occupation of $d_{xy}$ orbital through the crystal-field excitation cannot be neglected.
	Therefore, the orbital doublet is inadequate for constructing the low-energy effective model for this case.
	In contrast, the crystal-field excitation in $d^2$ configuration is suppressed by the interaction $U$ in the triangular lattice.
	While, for the case $t_{\mu\nu}\ll\square$, 
	we follow the procedure described in Appendix~\ref{app:Heff}.
	It is straightforward to show that the leading order Hamiltonian takes the following form  
	\begin{equation}
	\mathscr{H}^\text{eff}_\square = J_z \sum_{\langle i j \rangle} \tau_i^z\tau_j^z
	\end{equation}
	with the antiferro-orbital Ising coupling $J_z=2t^2_\sigma/U$ and 
	the pseudospin $\tau^z=\left(d^\dagger_{x^2+y^2}d_{x^2+y^2}-d^\dagger_{x^2-y^2}d_{x^2-y^2}\right)/2$.
	The antiferro-orbital coupling favors N\'eel ordering in the square lattice.
	Due to the extra constraint $t_{\sigma}\ll\square$, 
	it may require extremely low temperatures to experimentally detect the orbital ordering through the time-of-flight interference.  
	
	\section{Summary}
	\label{sec:summary}
	
	In conclusion, we have shown that the degeneracy of $d$ orbitals is lifted
	in both triangular and square optical lattices by a perturbative treatment.
	In particular, the selection rule is invoked in determining the symmetry reduction 
	from the $z$-axis rotation symmetry of harmonic oscillator approximation 
	to the discrete point group symmetry of optical potential. 
	We emphasize that our theory can be easily generalized 
	to the superstructured optical lattices, such as checkerboard lattice, 
	and is capable of predicting the orbital degeneracy from symmetry aspects. 
	Therefore our theory has potential applications in the quantum material design of optical lattices.
	Our work shall attract more experimental efforts in engineering $d$ orbitals,
	and may open fascinating new ground 
	for the quantum simulation of strongly correlated $d$-orbital physics in optical lattices.
	
	\section{Acknowledgments}
	
	We thank Haiwen Liu, Xiongjun Liu, Congjun Wu, and Hongyu Yang for helpful discussions. 
	This work is supported by 
	the National Natural Science Foundation of China under Grants No. 11704338, No. 11534001, and No. 11504008, 
	and the National Basic Research Program of China under Grant No. 2015CB921102. 
	
	\appendix
	\section{Algebraic Solutions of an Isotropic Two-Dimensional Harmonic Oscillator}
	\label{app:HO}	
	
	\begin{table*}
		\caption{\label{tab:eigfun-s}
			Eigenfunctions $\Psi_{\left[n,m\right]}\left(\bm{r}\right)$ of the 2D isotropic harmonic oscillator for $n=\{0,1,2\}$.}
		\begin{ruledtabular}
			\begin{tabular}{ccccc}
				$n \equiv n_+ + n_-$	&	$m \equiv n_+ - n_-$ &	$ n_+ $	&	$ n_-$	& 	$\Psi_{\left[n,m\right]}\left(\bm{r}\right)\equiv R_{\left[n,m\right]}\left(r\right)\exp\left[im\theta\right]$	\\
				\hline
				$n=0$	&	$m=0$	&	$n_+ = 0$	&	$n_- = 0$	&	$\Psi_{\left[0,0\right]}\left(\bm{r}\right)=\frac{\beta}{\sqrt{\pi}}\exp\left[-\frac{\beta^2 r^2}{2}\right]$\\
				\hline
				\multirow{2}{*}{$n=1$}  &	$m=+1$	&	$n_+ = 1$	&	$n_- = 0$	&	$\Psi_{\left[1,+1\right]}\left(\bm{r}\right)=\frac{\beta^2}{\sqrt{\pi}}r\exp\left[-\frac{\beta^2 r^2}{2}\right]\exp\left[+i\theta\right]$\\
				&	$m=-1$	&	$n_+ = 0$	&	$n_- = 1$	&	$\Psi_{\left[1,-1\right]}\left(\bm{r}\right)=\frac{\beta^2}{\sqrt{\pi}}r\exp\left[-\frac{\beta^2 r^2}{2}\right]\exp\left[-i\theta\right]$\\
				\hline
				\multirow{3}{*}{$n=2$}  &	$m=+2$	&	$n_+ = 2$	&	$n_- = 0$	&	$\Psi_{\left[2,+2\right]}\left(\bm{r}\right)=\frac{\beta^3}{\sqrt{2\pi}}r^2\exp\left[-\frac{\beta^2 r^2}{2}\right]\exp\left[+2i\theta\right]$\\
				&	$m=0$	&	$n_+ = 1$	&	$n_- = 1$	&	$\Psi_{\left[2,0\right]}\left(\bm{r}\right)=\frac{\beta}{\sqrt{\pi}}\left[\left(\beta r\right)^2-1\right]\exp\left[-\frac{\beta^2 r^2}{2}\right]$\\		
				&	$m=-2$	&	$n_+ = 0$	&	$n_- = 2$	&	$\Psi_{\left[2,-2\right]}\left(\bm{r}\right)=\frac{\beta^3}{\sqrt{2\pi}}r^2\exp\left[-\frac{\beta^2 r^2}{2}\right]\exp\left[-2i\theta\right]$\\			  						
				
			\end{tabular}
		\end{ruledtabular}
	\end{table*}
	
	We will derive the algebraic solutions of an isotropic 2D Harmonic oscillator that is described by the following Hamiltonian
	\begin{equation}
	\mathscr{H}_\text{HO}=\frac{\hat{\bm{p}}^2}{2M}+\frac{1}{2}M\omega^2r^2 \nonumber
	\end{equation}
	where $M$ is the mass of atoms trapped in the quantum well and $\omega$ is the harmonic frequency. 
	The isotropic 2D Harmonic oscillator can split into two 1D uncoupled oscillators in $\mu=x,y$ directions
	\begin{equation}
	\mathscr{H}_\mu=\frac{\hat{p}_\mu^2}{2M}+\frac{1}{2}M\omega^2\mu^2. \nonumber
	\end{equation}
	Let us first introduce the lowering and raising operators for the 1D harmonic oscillators
	\begin{eqnarray}
	a_\mu&=&\frac{1}{\sqrt{2}}\left(\beta \mu+i\frac{\hat{p}_\mu}{\beta\hbar}\right) \nonumber \\ 
	a_\mu^\dagger&=&\frac{1}{\sqrt{2}}\left(\beta \mu-i\frac{\hat{p}_\mu}{\beta\hbar}\right) \nonumber
	\end{eqnarray}
	with $\beta\equiv\sqrt{\frac{M\omega}{\hbar}}$. 
	In terms of number operators $\hat{n}_\mu=a_\mu^\dagger a_\mu$, the Hamiltonian of 2D oscillator can be rewritten as $\mathscr{H}_\text{HO}=\hbar\omega\left(\hat{n}_x+\hat{n}_y+1\right)$.
	Thus, the eigenfunctions $\psi_{\left[n_x,n_y\right]}\left(\bm{r}\right)$ of 2D oscillator, 
	corresponding to the energy $E=\hbar\omega\left(n_x+n_y+1\right)$, are characterized 
	by 1D harmonic oscillator quanta $n_\mu$ in $\mu=x,y$ directions. 
	Since the isotropic 2D Harmonic oscillator is invariant under rotation about the $z$-axis, the Hamiltonian $\mathscr{H}_\text{HO}$ should commute with the operator 
	$\hat{L}_z=x\hat{p}_y-y\hat{p}_x$ of infinitesimal rotation about $z$-axis, 
	{\it i.e.} the $z$-component angular momentum operator.
	In the following, we shall seek for a basis of eigenfunctions common to both $\mathscr{H}_\text{HO}$ and $\hat{L}_z$.
	To take better advantage of the continuous rotation symmetry, we introduce the chiral operators as follows 
	\begin{eqnarray}
	a_\pm^\dagger &=& \frac{1}{\sqrt{2}}\left(a_x^\dagger \pm ia_y^\dagger\right). \nonumber
	\end{eqnarray}	
	It is easy to verify that the only non-zero commutators between chiral operators are 
	$\left[a_+,a_+^\dagger\right]=\left[a_-,a_-^\dagger\right]=1$. 
	The corresponding number operators $\hat{n}_\pm=a_\pm^\dagger a_\pm$ 
	count the number of right($+$) and left($-$) circular quanta.
	With this definition, the Hamiltonian can be rewritten as $\mathscr{H}_\text{HO} = \hbar\omega\left(\hat{n}_+ + \hat{n}_-+1\right) \equiv \hbar\omega \left(\hat{n}+1\right)$
	with $\hat{n}\equiv\hat{n}_+ + \hat{n}_-$ being the total quanta operator.
	In addition, the $z$-component angular momentum operator can also be rewritten as  
	$\hat{L}_z = \hbar\left(\hat{n}_+ - \hat{n}_-\right) \equiv \hbar \hat{m}$ with $\hat{m}\equiv\hat{n}_+ - \hat{n}_-$. 
	Therefore, the eigenfunctions of $\mathscr{H}_\text{HO}$ can be characterized by either $\left[n_+,n_-\right]$ or $\left[n,m\right]$.
	The ground state $\Psi_{\left[n=n_++n_-=0,m=n_+-n_-=0\right]}\left(\bm{r}\right)$ contains 
	no right ($n_+=0$) and left ($n_-=0$) circular quanta and is identical to $\psi_{\left[n_x=0,n_y=0\right]}\left(\bm{r}\right)$ up to a phase.
	The eigenfunctions of excited states can be evaluated by applying the chiral operators $a_\pm^\dagger$ to the ground state 
	\begin{eqnarray}
	\Psi_{\left[n=n_+ + n_-,m=n_+ - n_-\right]}\left(\bm{r}\right) 
	=\frac{\left(a_+^\dagger\right)^{n_+}\left(a_-^\dagger\right)^{n_-}}{\sqrt{n_+! n_-!}}\Psi_{\left[0,0\right]}\left(\bm{r}\right). \nonumber
	\end{eqnarray}
	The explicit forms of eigenfunctions $\Psi_{\left[n,m\right]}\left(\bm{r}\right)$ for $n=\{0,1,2\}$ are listed in Table~\ref{tab:eigfun-s}.

	\section{Haldane Pseudopotential Descriptions of Interacting Hamiltonian}	
	\label{app:Haldane}	
	
	\begin{table*}
		\caption{\label{tab:chi-s} 
			Wave functions $\chi_{m_1m_2}^\pm\left(\bm{r}_+\right)$
			describe the center-of mass motion of two-particle states in Eq.~(\ref{eq:tps-s}).}
		\begin{ruledtabular}
			\begin{tabular}{cccc}
				&\multicolumn{3}{c}{$\chi_{m_1m_2}^+\left(\bm{r}_+\right)$}\\
				&	$m_2=-2$	&	$m_2=0$	&	$m_2=+2$	\\ \hline
				$m_1=-2$	
				&	$0$	
				&	$-\frac{\beta^6}{2\sqrt{2}\pi}r^3_+\exp\left[-3i\theta_+\right]$	
				&	$-\frac{\beta^6}{2\pi}r^3_+\exp\left[-i\theta_+\right]$	\\
				$m_1=0$	
				&	$\frac{\beta^6}{2\sqrt{2}\pi}r^3_+\exp\left[-3i\theta_+\right]$	
				&	$0$	
				&	$-\frac{\beta^4}{\sqrt{2}\pi}r_+\left(\frac{1}{2}\beta^2r^2_+-1\right)\exp\left[i\theta_+\right]$	\\	
				$m_1=+2$	
				&	$\frac{\beta^6}{2\pi}r^3_+\exp\left[-i\theta_+\right]$	
				&	$\frac{\beta^4}{\sqrt{2}\pi}r_+\left(\frac{1}{2}\beta^2r^2_+-1\right)\exp\left[i\theta_+\right]$	
				&	$0$	\\ \hline\hline	
				&\multicolumn{3}{c}{$\chi_{m_1m_2}^-\left(\bm{r}_+\right)$}\\
				&	$m_2=-2$	&	$m_2=0$	&	$m_2=+2$	\\ \hline
				$m_1=-2$	
				&	$0$	
				&	$\frac{\beta^4}{\sqrt{2}\pi}r_+\left(\frac{1}{2}\beta^2r^2_+-1\right)\exp\left[-i\theta_+\right]$	
				&	$\frac{\beta^6}{2\pi}r^3_+\exp\left[i\theta_+\right]$	\\
				$m_1=0$	
				&	$-\frac{\beta^4}{\sqrt{2}\pi}r_+\left(\frac{1}{2}\beta^2r^2_+-1\right)\exp\left[-i\theta_+\right]$	
				&	$0$	
				&	$\frac{\beta^6}{2\sqrt{2}\pi}r_+^3\exp\left[3i\theta_+\right]$	\\	
				$m_1=+2$	
				&	$-\frac{\beta^6}{2\pi}r^3_+\exp\left[i\theta_+\right]$	
				&	$-\frac{\beta^6}{2\sqrt{2}\pi}r_+^3\exp\left[3i\theta_+\right]$	
				&	$0$	\\					
			\end{tabular}
		\end{ruledtabular}
	\end{table*}
	
	The central interaction potential $\hat{U}\left(r\right)$ 
	that depends only on the relative coordinate $r$ between particle pairs
	can be described by a set of Haldane pseudopotentials $v_{m}$~\cite{Haldane90}.
	The potentials $v_m$ are obtained from the decomposition of two-particle states 
	into the states with relative angular momentum $m$.
	According to the Fermi (Bose) statistics, the many-particle state of fermions (bosons) 
	upon interchanging two particles is antisymmetric (symmetric), which requires that $m$ is odd (even).	
	For the present case of spinless fermions with short-range interaction, 
	we restrict the relative motion of two-particle states in the lowest odd angular momentum $m=\pm1$, 
	corresponding to the $p$-wave channel.
	Specifically, the two-particle state is factorized into two decoupled wave functions 
	that describe the center-of-mass ($\bm{r}_+\equiv\frac{1}{2}\left(\bm{r}_1+\bm{r}_2\right)$) motion 
	and the relative ($\bm{r}_-=\bm{r}_1-\bm{r}_2$) motion
	\begin{eqnarray}
	&&\Psi_{n=2,m_1}\left(\bm{r}_1\right)\Psi_{n=2,m_2}\left(\bm{r}_2\right) \approx
	r_-\exp\left[-\beta^2\left(r_+^2+\frac{r_-^2}{4}\right)\right] \nonumber\\
	&&\times\left\{\chi_{m_1m_2}^+\left(\bm{r_+}\right)\exp\left[i\theta_-\right]\right.
	+\left.\chi_{m_1m_2}^-\left(\bm{r_+}\right)\exp\left[-i\theta_-\right]\right\}, 
	\label{eq:tps-s}
	\end{eqnarray}
	where $\chi_{m_1m_2}^\pm\left(\bm{r}_+\right)$ are listed in Table~\ref{tab:chi-s}.
	In Eq.~(\ref{eq:tps-s}), we neglect the high-order terms in $r_-$ and keep the linear terms in the brace, 
	which corresponds to the short-range components of the interaction.
	Such an approximation is valid when the effective range of interaction 
	is much shorter than the characteristic length of 2D harmonic oscillator.
	It is straightforward to show that the interacting Hamiltonian takes the following form 
	\begin{equation}
	\mathscr{H}_\text{I} =\frac{1}{2}U_{m_1m_2m_3m_4}d^\dagger_{m_1}d^\dagger_{m_2}d_{m_3}d_{m_4} \nonumber
	\end{equation}
	with the interaction matrix
	\begin{eqnarray}
	&U&_{m_1m_2m_3m_4}\equiv 
	\int d^2\bm{r}_+ \frac{1}{\beta^4}\left[ v_{+1}\chi^{+*}_{m_2m_1}\left(\bm{r}_+\right) \chi^+_{m_3m_4}\left(\bm{r}_+\right) \right.\nonumber\\
	&+& \left. v_{-1}\chi^{-*}_{m_2m_1}\left(\bm{r}_+\right) \chi^-_{m_3m_4}\left(\bm{r}_+\right) \right]
	\exp\left[-2\beta^2r^2_+\right]
	\label{eq:int-s}
	\end{eqnarray}
	and the Haldane pseudopotentials 
	$v_{\pm1}=\beta^4 \int d^2\bm{r}_- r_- \exp\left[-\beta^2\frac{r_-^2}{4}\right] \hat{U}\left(r_-\right) r_- \exp\left[-\beta^2\frac{r_-^2}{4}\right]\equiv v$.
	A little algebra on the integral of Eq.~(\ref{eq:int-s}) over the center-of-mass coordinates $\bm{r}_+$ and a unitary basis transformation lead to the following Hamiltonian
	\begin{equation}
	\mathscr{H}_\text{I} = \frac{3v}{16\pi} \left[\left(\hat{n}_{x^2-y^2}+\hat{n}_{xy}\right)\hat{n}_{x^2+y^2}
	+2\hat{n}_{x^2-y^2}\hat{n}_{xy}\right]. \nonumber
	\end{equation}
	
	\section{The Derivation of Orbital Superexchange Hamiltonian $\mathscr{H}_\triangle^\text{eff}$}
	\label{app:Heff}	
	
	\begin{table*}
		\caption{\label{tab:gamma-s} 
			Eigenenergy $E_{\Gamma_n^i}$ and eigenstates $\Gamma_n^i$ of local Hamiltonian $\mathscr{H}_\triangle^\text{L}$ for $d^{n=1,2,3}$ configurations. 
			$\left|\text{vac}\right\rangle$ is the vacuum state.}
		\begin{ruledtabular}
			\begin{tabular}{cccccccc}
				&\multicolumn{3}{c}{$d^1$ configuration}&\multicolumn{3}{c}{$d^2$ configuration}&$d^3$ configuration\\ \hline
				$i$	
				&	$1$	&	$2$	&	$3$
				&	$1$	&	$2$	& 	$3$	
				&	$1$	\\ 
				$E_{\Gamma_n^i}$	
				&	$\Delta$	&	0	&	$\Delta$
				&	$U+\Delta$	&	$2U+2\Delta$	& 	$U+\Delta$	
				&	$4U+2\Delta$	\\ 	
				$\left|\Gamma_n^i\right\rangle$	
				&	$d_{xy}^\dagger\left|\text{vac}\right\rangle$	
				&	$d_{x^2+y^2}^\dagger\left|\text{vac}\right\rangle$	
				&	$d_{x^2-y^2}^\dagger\left|\text{vac}\right\rangle$
				&	$d_{xy}^\dagger d_{x^2+y^2}^\dagger\left|\text{vac}\right\rangle$	
				&	$d_{xy}^\dagger d_{x^2-y^2}^\dagger\left|\text{vac}\right\rangle$
				& 	$d_{x^2+y^2}^\dagger d_{x^2-y^2}^\dagger\left|\text{vac}\right\rangle$	
				&	$d_{xy}^\dagger d_{x^2+y^2}^\dagger d_{x^2-y^2}^\dagger\left|\text{vac}\right\rangle$	\\ 				
				
			\end{tabular}
		\end{ruledtabular}
	\end{table*}
	
	To derive the effective low-energy Hamiltonian, we first diagonalize the local on-site Hamiltonian as follow 
	\begin{eqnarray}
	\mathscr{H}_\triangle^\text{L}&\equiv& \mathscr{H}_\triangle + \mathscr{H}_\text{I}
	= \triangle \left(d_{x^2-y^2}^\dagger d_{x^2-y^2} + d_{xy}^\dagger d_{xy}\right) \nonumber\\
	&+& U\left[\left(\hat{n}_{x^2-y^2}+\hat{n}_{xy}\right)\hat{n}_{x^2+y^2}+2\hat{n}_{x^2-y^2}\hat{n}_{xy}\right] \nonumber\\
	&=&\sum_{\Gamma_n}E_{\Gamma_n^i}\left|\Gamma_n^i\right\rangle\left\langle\Gamma_n^i\right|. \nonumber
	\end{eqnarray}
	where $\Gamma_n^i$ is the $i$-th eigenstate of $d^n$ configuration with eigenenergy $E_{\Gamma_n}^i$.
	The eigenstates $\Gamma_n^i$ and eigenenergies $E_{\Gamma_n^i}$ for $d^{n=1,2,3}$ configurations 
	are listed in Table.~\ref{tab:gamma-s}. 
	In the large-$U$ limit, the ground state of $d^2$ configuration with energy $U+\Delta$ is 
	an orbital doublet $\Gamma_2^{1,3}$ with one fermion occupying $d_{x^2+y^2}$ and the other on occupying either $d_{xy}$ or $d_{x^2-y^2}$ orbital.
	Note that the doublet $\Gamma_2^{1,3}$ is well separated from the excited state $\Gamma_2^2$ by the energy gap $U+\Delta$. 
	Therefore, in the large-$U$ limit, it is reasonable to construct an effective model based on the doublet $\Gamma_2^{1,3}$ with the degenerate perturbation theory. 
	For convenience, we introduce the pseudospin operators  
	$\{\tau^+,\tau^-\} \equiv \{d^\dagger_{x^2-y^2}d_{xy}\hat{n}_{x^2+y^2},d^\dagger_{xy}d_{x^2-y^2}\hat{n}_{x^2+y^2}\}$,
	which flip the states of orbital doublet. The $z$ component of pseudospin $\bm{\tau}$-vector follows through the spin-$1/2$ angular-momentum algebra $\tau^z=\left[\tau^+,\tau^-\right]$. 
	Unlike for a spin system, the charge excitation 
	$\left(d^2\right)_i\left(d^2\right)_j\rightleftharpoons\left(d^3\right)_i\left(d^1\right)_j$,
	associated with the hopping process $t_{\mu\nu}d^\dagger_{i\mu}d_{j\nu}$, is directional dependent.
	It originates from the fact that the hopping process is anisotropic due to the spatial orientation of $d$ orbitals. 
	Let us first derive the superexchange interaction along $\bm{e}_1$ bonds as shown in Fig.~\ref{fig:triangle} (a) of main text.
	Employing the second-order perturbation theory~\cite{Kuklov03}, the matrix form of superexchang interaction is given by
	\begin{eqnarray}
	\left(J\right)_{kl;k^\prime l^\prime}=
	&-&\sum_{pq}\frac{1}{E_{\Gamma_{3}^p}+E_{\Gamma_1^q}-2\left(U+\Delta\right)} \nonumber\\
	&\times&\left\langle
	\underset{i\text{-site}}{\Gamma_2^k}
	\underset{j\text{-site}}{\Gamma_2^l}\right|
	\sum_{\mu\nu}t_{\mu\nu}^*d^\dagger_{j\nu}d_{i\mu}
	\left| \underset{i\text{-site}}{\Gamma_3^p}
	\underset{j\text{-site}}{\Gamma_1^q} 
	\right\rangle \nonumber\\
	&\times&\left\langle
	\underset{i\text{-site}}{\Gamma_3^p}
	\underset{j\text{-site}}{\Gamma_1^q}\right|
	\sum_{\mu^\prime\nu^\prime}t_{\mu^\prime\nu^\prime}d_{i\mu^\prime}^\dagger d_{j\nu^\prime}
	\left| \underset{i\text{-site}}{\Gamma_2^{k^\prime}}
	\underset{j\text{-site}}{\Gamma_2^{l^\prime}}
	\right\rangle \nonumber\\
	&+&{i\leftrightarrow j}. \nonumber
	\end{eqnarray}
	A lengthy but straightforward algebra on the summation of all bonds along the $\bm{e}_1$ vector leads to
	\begin{equation}
	\mathscr{H}_\triangle^{\bm{e}_1} = 
	J_\text{C} \sum_i \tau^z_{i}\tau^z_{i\pm\bm{e}_1} 
	+ J_\text{H}\sum_i\bm{\tau}_i\cdot\bm{\tau}_{i\pm\bm{e}_1}, \nonumber
	\end{equation}
	where 
	\begin{eqnarray}
	J_\text{C}&=&\left(t_{x^2-y^2,x^2-y^2}-t_{xy,xy}\right)^2/2U, \nonumber\\
	J_\text{H}&=&t_{xy,xy}t_{x^2-y^2,x^2-y^2}/U. \nonumber
	\end{eqnarray}
	The hopping term $t_{xy,xy}$ ($t_{x^2-y^2,x^2-y^2}$) denotes the intra-orbital hopping integral of $d_{xy}$ ($d_{x^2-y^2}$) orbital along the bond vector $\bm{e}_1$. 
	Note that the bond vector $\bm{e}_1$ lies in the nodal plane of $d_{xy}$ orbital and thus $t_{xy,xy}$ can be labeled by $\pi$-bonding $t_{\pi}$.
	While, the bonding state $t_{x^2-y^2,x^2-y^2}$ is symmetrical with respect to a $\pi$ rotation about the bond vector $\bm{e}_1$ and thus is labeled by $\sigma$-bonding $t_{\sigma}$.
	Having derived the superexchange model $\mathscr{H}_\triangle^{\bm{e}_1}$ along bond vector $\bm{e}_1$,
	the corresponding superexchange Hamiltonian $\mathscr{H}_\triangle^{\bm{e}_{2,3}}$ has exactly the same form
	with $\mathscr{H}_\triangle^{\bm{e}_{1}}$ 
	if the pseudospin operators $\bm{\tau}$ are defined in the local coordinate.
	In the local coordinate,  the local $x$ axis is defined  along the $\bm{e}_{2,3}$ bond vector.
	Thus, the connection between the local and global coordinates (the global $x$ axis along $\bm{e}_1$ bond vector) 
	is linked by a rotation of $\theta=\frac{2\pi}{3},\frac{4\pi}{3}$ about $z$ axis,
	corresponding to the $\bm{e}_2,\bm{e}_3$ bonds, respectively.
	The $d$-orbital wave functions transfrom under the rotation as 
	\begin{eqnarray}
	d_{x^2-y^2} &\to& \cos\left[2\theta\right] d_{x^2-y^2} - \sin\left[2\theta\right] d_{xy}, \nonumber\\
	d_{xy} &\to& \sin\left[2\theta\right]d_{x^2-y^2} + \cos\left[2\theta\right] d_{xy}, \nonumber \\
	d_{x^2+y^2} &\to& d_{x^2+y^2}. \nonumber
	\end{eqnarray} 	
	Accordingly, the pseudospin operators $\bm{\tau}$ transform as follow
	\begin{eqnarray}
	\tau^z &\to& \sin\left[4\theta\right]\tau^x+\cos\left[4\theta\right]\tau^z, \nonumber\\
	\tau^x &\to& \cos\left[4\theta\right]\tau^x-\sin\left[4\theta\right]\tau^z, \nonumber\\
	\tau^y &\to& \tau^y. \nonumber
	\end{eqnarray}	
	The pseudospin vector $\bm{\tau}$ is rotated by $4\theta$ about its $y$ axis in the pseudospin space.
	It is now straightforward to obtain the Hamiltonian $\mathscr{H}_\triangle^{\bm{e}_{2,3}}$ by replacing the pseudospin $\bm{\tau}$ in $\mathscr{H}_\triangle^{\bm{e}_1}$. 
	Finally, the total superexchange Hamiltonian takes the form 
	\begin{equation}
	\mathscr{H}_\triangle^\text{eff}\equiv 
	\sum_{i=1}^{3}\mathscr{H}_\triangle^{\bm{e}_i}
	=J_\text{C}\sum_{ i\gamma\eta }
	\tau_i^\gamma \tau_{i+\eta\bm{e}_\gamma}^\gamma
	+J_\text{H}\sum_{ i\gamma\eta } \bm{\tau}_i\cdot\bm{\tau}_{i+\eta\bm{e}_\gamma} \nonumber
	\end{equation} 
	with
	\begin{eqnarray}
	&\tau^\gamma=\tau^z\cos\left[4\theta_\gamma\right]+\tau^x\sin\left[4\theta_\gamma\right],
	\bm{e}_\gamma=\hat{x}\cos\theta_\gamma+\hat{y}\sin\theta_\gamma,& \nonumber \\
	&\{\theta_1,\theta_2,\theta_3\}=\{0,\frac{2\pi}{3},\frac{4\pi}{3}\},\eta=\pm 1.& \nonumber 
	\end{eqnarray}	
	Thus, the effective Hamiltonian is described by the Heisenberg-Compass model.


\begin{thebibliography}{99}
\bibitem{dx2y2} The Wannier function $\Psi_{\left[n=2,m=0\right]}$ is defined as $s_{r^2-1}$ orbital in Ref.~\cite{Kock16}. Here we defined as the $d_{x^2+y^2}$ orbital 
		since its orientational anisotropy is identical to that of $d_{x^2+y^2}$ orbital in solid-state systems.
\end{thebibliography}
\end{document}